%% file: article.tex
\documentclass[doublecol]{epl2} 
\usepackage{graphicx,bm,amsmath,amsfonts}

\input{commands.tex}

\title{
 Statistical mechanical assessment of a reconstruction limit
 of compressed sensing: Toward theoretical analysis of correlated signals
}
\shorttitle{
 Statistical Mechanical Assesment of
 a Reconstruction Limit of Compressed Sensing
}

\author{Koujin Takeda \and Yoshiyuki Kabashima}
\institute{
 Department of Computational Intelligence and Systems Science,
 Tokyo Institute of Technology,
 Yokohama 226-8502, Japan
}
\shortauthor{K. Takeda \etal}


\abstract{
 We provide a scheme for exploring the reconstruction
 limits of compressed sensing by minimizing the general
 cost function under the random measurement constraints
 for generic correlated signal sources. Our scheme is
 based on the statistical mechanical replica method for
 dealing with random systems. As a simple but non-trivial
 example, we apply the scheme to a sparse autoregressive model,
 where the first differences in the input signals of the correlated
 time series are sparse, and evaluate the critical compression
 rate for a perfect reconstruction. The results are in good
 agreement with a numerical experiment for a signal reconstruction.
}

\begin{document}

\maketitle

\section{Introduction}
 Compressed sensing (CS) is a novel technique for data compression and
 has been drawing a lot of attention recently from the viewpoints of both
 theory and application. The key idea behind CS is to utilize the
 sparsity of the original input signals as the prior knowledge during the
 signal reconstruction stage, which can significantly reduce the number
 of signal measurements required for a perfect reconstruction. This setup
 is realistic because we often have to face situations where we have to
 handle sparse signals in the real world. A lot of effort has been paid
 and significant progress has been made in investigating the
 properties of CS \cite{CRT2006,D2006-1,CT2006}. After the pioneering
 works, contribution to CS problem from statistical
 mechanics analysis is now growing rapidly
 \cite{DMM,KWT,RFG,GS,BM,TK2010,TR}.

 The measurement process of CS is summarized in the following linear equation: 
\begin{equation} 
 \bm y = \bm F \bm x^{0}. 
\end{equation} 
 The vectors and matrices are denoted in bold in this article.
 The input signal vector $\bm x^{0}$ is $N$-dimensional and the
 compressed signal vector $\bm y \in \Real^{P}$ is $P$-dimensional.
 $\bm F$ is a $P$-by-$N$ compression matrix. In this article,
 we particularly focus on random measurements, in which each
 $\bm F$, $F_{\mu i}$ entry is independently and identically
 distributed (i.i.d.) from a Gaussian distribution of the zero mean
 and variance $N^{-1}$. The compression rate is defined by
 $\alpha \equiv P/N< 1 $.

 In earlier theoretical studies, the critical compression rate
 $\alpha_c$ for perfectly reconstructing $\bm x^0$ from $\bm y$
 has been actively assessed for various reconstruction schemes
 under the assumption that the input signal vector $\bm x^0$ is
 sparsely modeled by the distribution, 
\begin{equation} 
\label{eq:P0dist} 
 P(x^0_i) = (1 - \rho) \delta(x^0_i)
 + \rho \tilde{P}(x_i^0),
\end{equation} 
 within the large system limit of $N,P \to \infty$ keeping
 $\alpha=P/N$ constant \cite{CRT2006,D2006-1,CT2006,KWT}. Here,
 $\tilde{P}(x_i^0)$ is a given probabilistic distribution and
 $\rho$ denotes the density of the non-zero elements. In particular,
 the assessment for the reconstruction scheme for minimizing the
 so-called $\ell_1$-norm 
\begin{equation} 
\label{L1reconstruction}
 {\rm minimize} \sum_{i} |x_i| {\ \ \ \ \rm subject \ to \ \ } \bm y(=\bm F 
 {\bm x}^0)= \bm F \bm x, 
\end{equation} 
 which is termed the $\ell_1$-norm reconstruction hereafter,
 has drawn a lot of attention because of its computational
 feasibility and robustness to measurement noise.
 In this regard, it may be surprising that a
 mathematically rigorous method of combinatorial geometry
 \cite{D2006-1} and the replica method for statistical mechanics
 \cite{KWT} provide an identical $\alpha_c$ value although the
 methodological equivalence between the two schemes has not really
 been clarified yet. In addition, the value of $\alpha_c$ seems rather
 {\em universal} \cite{KWT,DT2009b,DT2010}; $\alpha_c$ is unchanged
 as long as $\bm x^0$ follows (\ref{eq:P0dist}) and
 ${\bm F}^{\rm T} {\bm F}$, where ${\rm T}$ denotes the matrix
 transpose, {\em asymptotically} obeys a rotationally invariant
 ensemble. However, the necessary and sufficient condition for
 the universality is still also open. 

 The main purpose of this article is to offer a methodological basis
 for exploring this universality using the replica method.
 For this objective, we evaluate $\alpha_c$ for general
 {\em correlated} distributions of $\bm x^0$, where
 $P(\bm x^0)$ is a joint distribution with sparsity and
 not necessarily factorizable to each
 $P(x^0_i)$,
 and the reconstruction
 schemes provided as 
\begin{equation} 
\label{def:general}
 {\rm minimize}\ \ E (\bm x) \ \ {\rm subject \ to \ \ } \bm y(=\bm F 
 {\bm x}^0)= \bm F \bm x, 
\end{equation} 
 where $E(\bm x)$ is a generic cost function. For simplicity,
 we assume that each entry of $\bm F$, $F_{\mu i}$, is an i.i.d.
 Gaussian random number of the zero mean and variance $N^{-1}$.
 However, as shown later, situations in which $\bm x^0$ is expanded
 by the i.i.d. coefficients sampled from (\ref{eq:P0dist})
 using a certain basis $\bm S$ can be cast to those of the
 correlated $F_{\mu i}$ for i.i.d. signals sampled from
 (\ref{eq:P0dist}). Namely, our analysis practically covers
 {\em correlated compression matrices} as well \cite{TK2010}.

 In addition to the theoretical interest, exploring the above setting
 is also significant for practical relevance.
 In most real world problems, the signals may be redundant
 in an information theoretic sense, but are not necessarily
 expressed as sparse upon first sight. In addition, in order to
 appropriately deal with such real world signals,
 certain cost functions other than the na\"{i}ve $\ell_1$-norm
 of (\ref{L1reconstruction}), such as the {\em total variation (TV)}
 \cite{Rudin1992}, are widely used in practice for reconstructing
 signals. Our generic assumptions concerning the correlated
 distributions of the signal sources and cost functions for the
 signal reconstruction are intended to extend the analysis of
 the performance measure of compressed sensing, $\alpha_c$,
 for more practically plausible scenarios beyond the simple cases
 of i.i.d. sparse sources and component-wise cost functions. 

\section{Replica analysis: A general guideline} 
 Here we sketch an outline of our analysis. This analysis is similar
 to that of the recent study regarding CS for correlated compression
 matrices \cite{TK2010} and that of the correlated channel
 in wireless telecommunication systems \cite{TUK,HTK}.
 The technical details can be found in these references.

 Following the basic scenario in \cite{KWT}, let us define the key
 quantity for our analysis, which plays the role of free energy
 in statistical mechanics and represents the typical value (per element)
 of the minimized cost (\ref{def:general}) in the current context, 
\begin{eqnarray} 
\label{eq:freeenergy}
 C & \equiv &
 - \lim_{\beta \to \infty} \frac{1}{\beta N} 
 [ \ln Z(\beta,\bm y) ]_{\bm F,\bm x^0} \nonumber \\
 &=& - \lim_{\beta \rightarrow \infty} \lim_{n \rightarrow 0} 
 \frac{\partial}{\partial n} \lim_{N \rightarrow \infty}
 \frac{1}{\beta N} \ln [ Z^n (\beta,\bm y)]_{\bm F, \bm x^0}, 
\end{eqnarray} 
 where $Z(\beta,\bm y)\equiv \int d\bx \exp(-\beta E(\bx))
 \delta(\bm F(\bm x-\bm x^0))$ is the partition function and
 $[\cdots ]_X$ generally denotes the average with respect
 to random variable $X$. Taking the limit $\beta \to \infty$ works for  
 singling out the solution of (\ref{def:general}) in the partition function. 
 Unfortunately, assessing $[Z^n(\beta,\bm y)]_{\bm F,\bm x^0}$ for
 $\forall{n} \in \mR$ in (\ref{eq:freeenergy}) is technically difficult.
 For resolving this difficulty, we  evaluate
 analytical expressions of
 $[Z^n(\beta,\bm y)]_{\bm F,\bm x^0}$ with respect to 
 $\forall{n} \in \mN$ using the identity
\begin{eqnarray} 
\label{eq:ffff1}
 Z^{n}(\beta, \bm y)  =  \int \prod_{a=1}^{n} {d \bm x^a}
 \exp (- \beta E (\bm x^a) )
 \delta (\bm F (\bm x^a - \bm x^0)),
\end{eqnarray} 
 which is valid only for $n \in \mN$, and 
 employ the obtained expressions for assessment of 
 (\ref{eq:freeenergy}) assuming that they hold for
 $\forall{n} \in \mR$ as well.
 This is often termed
 the replica method as integration variables $\bm x^a$ $(a=1,2,\ldots,n)$
 in (\ref{eq:ffff1}) are regarded as $n$ ``replicas'' of the
 original state variable $\bx$.
 For this, we analytically calculate the average of the right hand side
 of (\ref{eq:ffff1}) employing the saddle-point method
 with respect to macroscopic variables 
 $q_{ab}=N^{-1} (\bm x^a)^{\rm T} \bm x^b$ and $m_a=N^{-1} (\bm x^0)^{\rm T}
 \bx^a$,
 which is justified as $P,N \gg 1$.  The intrinsic invariance of
 (\ref{eq:ffff1}) under any permutations of replica indices
 $a=1,2,\ldots,n$ leads to
 the replica symmetric (RS) ansatz, which means that the dominant saddle
 point also possesses this property as $q_{aa}=Q$, $q_{ab}=q$ $(a \ne b)$
 and $m_a=m$. 
 This reproduces the mathematically rigorous results for the
 basic model \cite{KWT}. Therefore, we here also adopt this ansatz, 
 validity of which
 will be checked later.  The saddle point solution obtained under the RS
 ansatz seems to hold for $n \in \mR$ as well. Employing this in the
 right hand side of (\ref{eq:freeenergy}) yields an expression 
\begin{eqnarray} 
\label{eq:Cpfinal}
 C &=& \hspace{-0.5cm} 
 \extr_{q,m,\chi,\widehat{Q},\widehat{m},\widehat{\chi}}
 \left( \left. \frac{ \alpha ( q -2m + u)}{2 \chi}
 + \left( \frac{\chi \widehat{\chi}}{2}
 - \frac{q \widehat{Q}}{2} + m \widehat{m} 
 \right) \right. \right. \nonumber \\
 && \hspace{1cm} + 
 \left. \left\{\int d {\bm x}^0 P({\bm x}^0) \int D \tilde{\bm z} 
 \phi( \bm \omega, \widehat{Q}) \right\} \right).
\end{eqnarray}
 Here $\bm \omega =(\omega_i)\equiv \widehat{m}\bm x^0
 + \sqrt{\widehat{\chi}} \bm z$,
 $\extr_{\Theta} \{ \cdots \}$ denotes the extremization of
 $\cdots$ with respect to $\Theta$, $P(\bm x^0)$ is the generic
 $N$-dimensional distribution of the original signal $\bx^0$, and $u =
 N^{-1} \int d {\bm x}^0 P({\bm x}^0) |{\bm x}^0 |^2$ denotes the second
 moment (per element) of the original signal. $D\tilde{\bm z}$ stands
 for the $N$-dimensional Gaussian measure $(2\pi)^{-N/2}\prod_{i=1}^N
 d\tilde{z}_i \exp \left (-\tilde{z}_i^2/2 \right )$. The function
 $\phi(\bm h, \widehat{Q})$ is defined by the minimization including
 the $N$ variables as
\begin{equation} 
\label{eq:defphi}
 \phi(\bm h,\widehat{Q}) \hspace{-0.1cm} \equiv \hspace{-0.1cm} \frac{1}{N} 
 \mathop{\rm min}_{\bm x} \left\{ \frac{\widehat{Q}}{2} \bm x^{\rm T}
 \bm x - \bm h^{\rm T} \bm x + E(\bm x) \right\}. 
\end{equation} 

 With regard to the final expressions (\ref{eq:Cpfinal}) and 
 (\ref{eq:defphi}), three points are worthwhile to note. First, 
 the right hand side of (\ref{eq:defphi}), in conjunction with substitution of 
 ${\bm h}={\bm \omega}$ and $\widehat{Q}$  as provided by (\ref{eq:Cpfinal}), 
 stands for the problem statistically equivalent to the original one 
 (\ref{def:general}). 
 This means that random constraints ${\bm y}={\bm F} \bm x$
 of (\ref{def:general}), 
 in which multiple variables are coupled with one another, can be 
 handled as a bunch of decoupled extra random costs 
 $(\widehat{Q}/2) x_i^2 - \omega_i x_i$ ($i=1,2,\ldots,N$)
 in the performance assessment of large systems. 
 Such correspondence is sometimes termed ``decoupling principle''
 in information theory literature  
 \cite{GuoVerdu2005}. 
 
 Second, the values of $q$ and $m$
 determined by the extremization condition of the right hand side of
 (\ref{eq:Cpfinal}) represent the typical values of the averages of
 $N^{-1} {\bm x}^{\rm T} \bm x$ and $N^{-1} ({\bm x^0})^{\rm T} \bm x$
 with respect to the uniform distribution of the solutions of
 (\ref{def:general}), respectively. If and only if the solutions
 typically accorded to $\bm x^0$ allowing negligible errors per
 component in $N \to \infty$, the solution for $q=m=u$ is
 thermodynamically dominant, implying that the reconstruction is
 typically successful. Therefore, one can characterize $\alpha_c$ as a
 transition condition at which the successful solution $q=m=u$ loses its
 thermodynamic dominance. 
 When $E(\bx)$ is convex downward, 
 which is often the case in practice, this can be examined
 by assessing the local stability of $q=m=u$ since
 (\ref{def:general}) is guaranteed to possess a unique solution. 
 It might also be noteworthy that our
 criterion for a successful reconstruction is different from that of
 earlier mathematical studies \cite{CRT2006,D2006-1,CT2006}
 in which no errors were permitted. 
 However, we expect that such differences are irrelevant in the
 $\alpha_c$ assessment as was the case for the basic problems of
 (\ref{eq:P0dist}) and (\ref{L1reconstruction}) \cite{KWT}. 

 The final point is the computational cost for carrying out the
 above assessment. Although the average with respect to $\bm F$ has
 already been analytically taken into account, those with respect to $\bm
 x^0$ and auxiliary random numbers $\tilde{\bm z}$ still remain in the
 expression (\ref{eq:Cpfinal}). In practice, this should be assessed
 using a Monte Carlo sampling method for sufficiently large $N$ and $P$,
 which in principle can offer arbitrarily accurate estimates of the
 averages in the large system limit $N,P \to \infty$ (under the
 assumption that a certain thermodynamic limit exists). Therefore, the
 computational cost for performing the Monte Carlo sampling practically
 determines the feasibility. There are two possible sources for the
 computational difficulty. The first one is the computational cost for
 generating $\bm x^0$ following $N$-dimensional distribution $P(\bm
 x^0)$, which generally grows exponentially with respect to
 $N$. However, when $\bm x^0$ can be expressed as $\bm x^0={\bm S} \bm
 x^\prime$, where ${\bm S}$ and $\bm x^\prime$ are a fixed matrix and a
 vector sampled from a computationally feasible distribution,
 respectively, generating $\bm x^0$ is not a crucial problem for
 standard computational resources to date. This is also the case for
 $\tilde{\bm z}$. The other difficulty could come out in numerically
 performing a minimization with respect to $\bm x$ in
 (\ref{eq:defphi}). However, when $E(\bm x)$ is convex, which we are
 assuming, the cost function on the right hand side of (\ref{eq:defphi})
 is guaranteed to be convex as well. This indicates that one can also
 avoid a computational explosion using various schemes known for convex
 optimization \cite{GB1,GB2} in assessing
 (\ref{eq:defphi}). Furthermore, when the variable dependence of $E(\bm
 x)$ is pictorially expressed as a graph free from cycle, one may be
 able to use more efficient algorithms for the minimization
 \cite{Pearl1988}. These imply that although performing the developed
 method is generally computationally difficult, it is still practically
 useful in the performance analysis for certain non-trivial classes
 of CS problems. In the next part, this is illustrated through
 application to time series data signals that are characterized by
 the sparsity concerning the difference between signals of successive times. 

\section{Application: A sparse autoregressive model}
\subsection{Model definition}
 For illustrating the utility of the developed scheme,
 we focus on the time series data signals generated
 from the use of the autoregression process of
 the first order with sparsity
 (sparse AR(1) model, denoted by SAR(1) in the following).
 A SAR(1) process is defined by the stochastic recurrence equation 
\begin{equation}
\label{SARdefinition}
 x^0_{i+1} = \left\{
\begin{array}{ccc}
 r x^0_{i} + \sqrt{1-r^2} \eta_{i} & {\rm with \ prob.} & \rho, \\
 x^0_{i} & {\rm with \ prob.} & 1-\rho, \\
\end{array}
 \right.
\end{equation}
 where $0 \le r,\rho \le 1$.
 We assume that random variable $\eta_i$ at each time $i$,
 including the first signal $x_1^0$, is independently drawn from
 the normal Gaussian distribution ${\cal N} (0,1)$. Equivalently,
 this process is represented by the conditional probability of
 the signal at time $i$ given a state at time $i-1$, $x_{i-1}^0$, as
\begin{eqnarray} 
\label{eq:ARdist}
 P(x^0_{i} | x^0_{i-1}) &=& (1 - \rho) \delta(x^0_{i} -
 x^0_{i-1}) \nonumber \\
 && \hspace{-1.5cm}
 + \frac{\rho}{\sqrt{2 \pi (1-r^2)}}
 \exp \left( - \frac{(x^{0}_{i} - r x^{0}_{i-1})^2}{2 (1-r^2)} \right).
\end{eqnarray}
 The CS of this process has already been investigated from algorithmic
 point of view \cite{SZ}. Here we address the critical compression rate
 $\alpha_c$ of the signals from this process by the replica analysis.
 Although for simplicity reasons we focus on SAR(1) in the current
 article, extending the following argument to that of the $k$-th order,
 SAR($k$), is straightforward. 

\begin{figure} 
\begin{picture}(0,120)
 \put(30,-10){\includegraphics[width=0.35\textwidth]{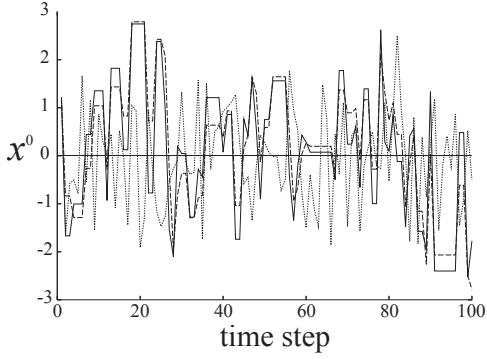}} 
\end{picture} 
\caption{Example of data signal from SAR(1).
 The cases of $\rho=0.5, r=0$ (solid), $\rho=0.5, r=0.5$ (broken),
 and $\rho=1, r=0$ (dotted) are shown here.}
\label{fig:figure1}
\end{figure}

 This model is considered as a special {\em Gaussian mixture transition
 distribution model} proposed by Le {\em et al.} for handling the
 non-Gaussian and nonlinear features of a time series in a unified
 framework \cite{Le1996,WongLi2000}. In (\ref{SARdefinition}) and
 (\ref{eq:ARdist}), $r$ represents a parameter of the autoregression
 satisfying $0 \le r \le 1$, while $\rho$ $(0 \le \rho \le 1)$ stands
 for a density parameter with respect to the difference in signals
 between successive times. An example of the signals from SAR(1) is
 depicted in figure \ref{fig:figure1}. For $\rho=1$ this process is
 reduced to a normal autoregressive model of the first order, and
 for $\rho<1$ the signal at time $i$ pauses for the same state as
 the one in the previous time step $i-1$ with a finite probability
 $1-\rho$. Therefore, SAR(1) of $\rho < 1$ typically generates a time
 series that has a lot more pausing states than usual autoregressive
 models. This property may be suitable for modeling various kinds
 of time series data such as acoustic signals \cite{Barber2006},
 the exploratory behavior of a house fly \cite{Takahashi2007},
 the financial time series \cite{Sazuka2003}, and more. 
 
 In SAR(1), the signal differences are sparse but the signals themselves
 are dense. This indicates that using the na\"{i}ve $\ell_1$-norm
 as a cost function for the signal reconstruction is not promising
 for improving the reconstruction performance. Instead, it may be
 reasonable to choose the cost function $E(\bm x)$ as $\ell_1$-norm
 for the signal differences, namely 
\begin{equation} 
\label{eq:diffL1norm}
 {\rm minimize} \sum_{i} | x_{i+1} - x_i |
 \ \ {\rm subject \ to \ \ } \bm y(=\bm F {\bm x}^0)= \bm F \bm x, 
\end{equation} 
 in terms of striking a balance between the statistical accordance
 to the original signals and computational feasibility. 

 Defining a vector of the signal differences as $x'_i = x_{i} - x_{i-1}$
 (and $x'_1=x_1$) formally converts (\ref{eq:diffL1norm}) into an
 expression of the na\"{i}ve $\ell_1$-norm reconstruction for
 $\bm x^\prime=(x_i^\prime)$ subject to the constraint
 $\bm y= \bm F^\prime \bm x^\prime$ offered by a modified compression matrix 
 $\bm F^\prime = \bm F \bm S,$
 where $\bm S=(S_{ij})$ is provided as
 $S_{ij}=1$ for $i \ge j$ and vanishes, otherwise. 
 Although $\bm F^\prime$ is also a certain random matrix,
 the ensemble of $(\bm F^\prime)^{\rm T} \bm F^\prime$ is no more
 rotationally invariant as
 $[ (\bm F^\prime)^{\rm T} \bm F^\prime]_{\bm F}=\bm S^{\rm T} \bm S$
 holds true. Therefore, one cannot apply the results of earlier studies
 for the basic settings \cite{CRT2006,D2006-1,CT2006,KWT} to the
 analysis of SAR(1), as was pointed out in \cite{TK2010}. 

\subsection{Saddle point equation and critical condition}
 Let us evaluate the critical reconstruction limit of SAR(1) by using
 the scheme developed in the preceding part for (\ref{eq:ARdist})
 and (\ref{eq:diffL1norm}).
 Note that (\ref{eq:diffL1norm})
 is described by a spin chain with random fields. Similar 
 problems have been analyzed in \cite{DH,WM}.
 Extremization of (\ref{eq:Cpfinal}), in
 conjunction with the substitution of
 $P(\bm x^0)=\prod_{i=1}^N P(x_i^0|x_{i-1}^0)$, where
 $P(x_1^0|x_0^0) = \exp(-x_1^2/2)/\sqrt{2\pi}$, yields a set of saddle
 point equations, as 
\begin{eqnarray} 
\label{eq:saddle}
 \widehat{Q} \hspace{-2mm} &=& \hspace{-2mm}
 \widehat{m} \ = \ \frac{\alpha}{\chi}, \hspace{2mm}
 \widehat{\chi}
 = \frac{\alpha(q - 2m + u)}{\chi^2}, \nonumber \\
 q \hspace{-2mm} &=& \hspace{-3mm}
 \int \prod_{i=1}^N D \tilde{z}_i d x_i^0 P(x_i^0 | x_{i-1}^0)
 \hspace{-1mm} \left ( \frac{1}{N} \sum_{j}
 \left( x_j^* (\bm \omega, \widehat{Q})
 \right)^{\! 2} \right)\hspace{-1mm},
 \nonumber \\
 m \hspace{-2mm} &=& \hspace{-3mm}
 \int \prod_{i=1}^N D \tilde{z}_i d x_i^0 P(x_i^0 | x_{i-1}^0) 
 \hspace{-1mm} \left ( \frac{1}{N} \sum_{j}
 x_j^0 x_j^* (\bm \omega, \widehat{Q})
 \right )\hspace{-1mm},
 \nonumber \\
 \chi \hspace{-2mm} &=& \hspace{-2mm}
 \frac{1}{\sqrt{\widehat{\chi}}}\int \prod_{i=1}^N D
 \tilde{z}_i d x_i^0 P(x_i^0 | x_{i-1}^0)
 \hspace{-0.5mm} \left( \hspace{-1mm} \frac{1}{N} 
 \sum_{j} \ \tilde{z}_j x_j^* (\bm \omega, \widehat{Q})
 \hspace{-0.5mm}\right)\hspace{-1mm},\nonumber \\
\end{eqnarray} 
 where $D z \equiv d z \exp \left( - z^2 / 2 \right) / \sqrt{2 \pi}$
 and the $N$-dimensional vector
 $\bm x^*=(x_j^*(\bm \omega, \widehat{Q}))$
 is determined by
\begin{eqnarray}
\label{eq:extremizex} 
 \frac{\partial}{\partial x_i^*}
 \phi(\bm \omega, \widehat{Q}) 
 &=& (\widehat{Q} x_i^* - \widehat{m} x_i^0 ) - \sqrt{\widehat{\chi}}
 \tilde{z}_i \nonumber \\
 && \hspace{-1.5cm} + {\rm sgn} (x_i^* - x_{i+1}^*)
 + {\rm sgn} (x_i^* - x_{i-1}^*) = 0
\end{eqnarray}
 $(i=1,2,\ldots,N)$, which corresponds to the minimization condition of
 $\phi(\bm \omega, \widehat{Q})$.
 ${\rm sgn}(x)=x/|x|$ for $x\ne 0$. 

 For a sufficiently large $\alpha$ given $r$ and $\rho$, the set
 of equations (\ref{eq:saddle}) allows for the following solution:
 $\chi \to +0$, $\widehat{Q}=\widehat{m} \to +\infty$, $Q=m \to u$
 and $\widehat{\chi} \sim O(1)$. This is because the third to
 fifth terms in (\ref{eq:extremizex}) are negligible compared to
 the first and second ones if $|x_j^* -x_j^0| \sim O(1)$ as
 $\widehat{Q}=\widehat{m} \to +\infty$ while $\widehat{\chi}$ is
 kept at $O(1)$, and therefore,
 $x_j^*(\bm \omega,\widehat{Q})
 \to x_j^0$ $(j=1,2,\ldots,N)$
 holds in (\ref{eq:extremizex}). This solution represents nothing
 but a successful reconstruction. 

 The critical reconstruction rate $\alpha_c$ is determined by
 the local instability condition of this solution, which is
 summarized as the condition for preventing the behavior of
 $\chi \to +0$. In order to accurately evaluate this, we pay attention
 to the infinitesimal differences between $x_i^*$ and $x_i^0$ by
 introducing the novel variables
 $\widehat{x}_i (\sqrt{\widehat{\chi} \bm z})
 \equiv \lim_{\chi \to +0} (\alpha/\chi) (x_i^*-x_i^0)$
 $(i=1,2,\ldots,N)$. Rewriting (\ref{eq:saddle}) using these variables
 within the limit of $\chi \to +0$ and exploring the local stability
 condition of $\chi \to +0$ yield a set of equations for determining
 the reconstruction limit $\alpha_c$,
\begin{eqnarray} 
\label{eq:finalresultAR}
 \widehat{\chi} \hspace{-3mm} &=& \hspace{-3mm} 
 \frac{1}{\alpha_c}\int \prod_{i=1}^N 
 D \tilde{z}_i d {x}_i^0 P({x}_i^0 | {x}_{i-1}^0)
 \! \left( \! \frac{1}{N}
 \sum_{j=1}^N
 (\widehat{x_j}(\sqrt{\widehat{\chi}}\tilde{\bm z}))^2
 \! \right) \!\!,
 \nonumber \\
 \alpha_c \!\!\!\!\! &=& \!\!\!\!\!\!
 \frac{1}{\sqrt{\widehat{\chi}}} \!
 \int \prod_{i=1}^N D 
 \tilde{z}_i d x_i^0 P({x}_i^0 | {x}_{i-1}^0)
 \! \left( \! \frac{1}{N}
 \sum_{j=1}^N \ \tilde{z}_j \widehat{x}_j
 (\sqrt{\widehat{\chi}}\tilde{\bm z}) \! \right) \!\!, \nonumber \\
 && 
 \hspace{-0.5cm} \widehat{x}_i
 - \sqrt{\widehat{\chi}} \tilde{z}_i
 + {\rm sgn} ( \epsilon(\widehat{x}_i-\widehat{x}_{i+1})+ x^0_i - x^0_{i+1})
 \nonumber \\
 && \hspace{1cm}
 + {\rm sgn} (\epsilon(\widehat{x}_i-\widehat{x}_{i-1})+x^0_i - x^0_{i-1})=0, 
\end{eqnarray} 
 where $\epsilon>0$ is a sufficiently small positive constant. 
 We also checked the local stability  
 against the disturbance that breaks the replica symmetry \cite{AT}, 
 which gives the stability condition as 
\begin{eqnarray}
\label{ATcond}
 &&\hspace{-0.7cm}\frac{\alpha}{\chi^2}
 \int \prod_{i=1}^N 
 D \tilde{z}_i d {x}_i^0 P({x}_i^0 | {x}_{i-1}^0)
 \! \left\{ \! \frac{1}{N} \sum_{j,k} \left(
 \frac{\partial x_j^* (\bm \omega, \widehat{Q} ))}
 {\partial \omega_k} \right)^{\!\!\! 2} \! \right\}
 \! < \! 1. \nonumber \\
 &&
\end{eqnarray}
 It may be noteworthy that this accords to that for the {\em dynamical 
 stability} of the successful solution $\bx_*=\bx^0$ concerning a 
 belief propagation based algorithm for solving 
 (\ref{eq:diffL1norm}). 
 A similar accordance has been   
 observed in another system before\cite{Kabashima2003}.  

 Unfortunately, the off-diagonal contributions 
 of $(\partial x_j^*(\bm \omega, \widehat{Q} )/\partial \omega_k)^2$ 
 $(j \ne k)$ always
 prevent the solution of (\ref{eq:finalresultAR}) from satisfying 
 (\ref{ATcond}). This implies the necessity of exploring 
 the replica symmetry breaking (RSB) solutions 
 for accurately assessing $\alpha_c$. 
 However,  we still speculate that the RS estimate 
 at least offers a fairly good approximation since the deviation from 
 the results of numerical experiments shown later is considerably small. 
 This speculation is also supported by the fact that the RS assessment 
 provides the correct estimate of $\alpha_c$ of the 
 $l_0$ recovery scheme for the basic model
 in spite that the RS solution is locally unstable for 
 the RSB disturbance \cite{KWT}. 

 In the evaluation of the reconstruction limit $\alpha_c$, multiple
 integrals in the first and the second equations in
 (\ref{eq:finalresultAR}) should be performed. This can be done in
 practice by using a Monte Carlo method. Particularly in the current
 case, this scheme works very efficiently because the subroutine for
 determining $\widehat{x}_i (\sqrt{\widehat{\chi}}\tilde{\bm z})$,
 which is expressed as the third equation,
 can be carried out by using only the $O(N)$ computational cost for a
 given pair of $\bm x^0$ and $\tilde{\bm z}$ with making use of the
 belief propagation (equivalently, transfer matrix method or dynamic
 programming) \cite{Pearl1988}. 

\subsection{Monte Carlo assessment of $\alpha_c$ and experimental validation}
\begin{figure} 
\begin{picture}(0,250)
 \put(30,125){\includegraphics[width=0.34\textwidth]{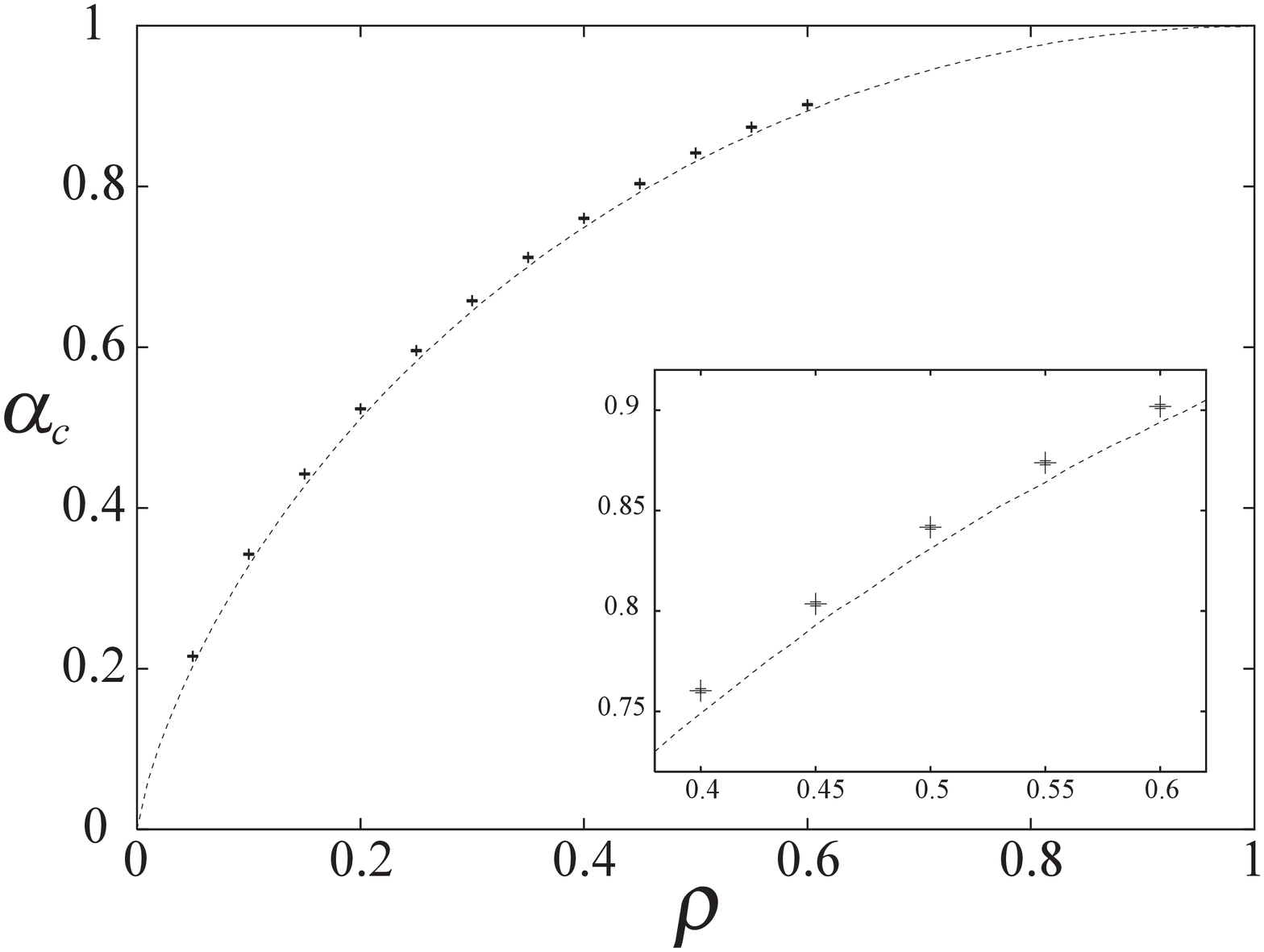}} 
 \put(30,-12){\includegraphics[width=0.34\textwidth]{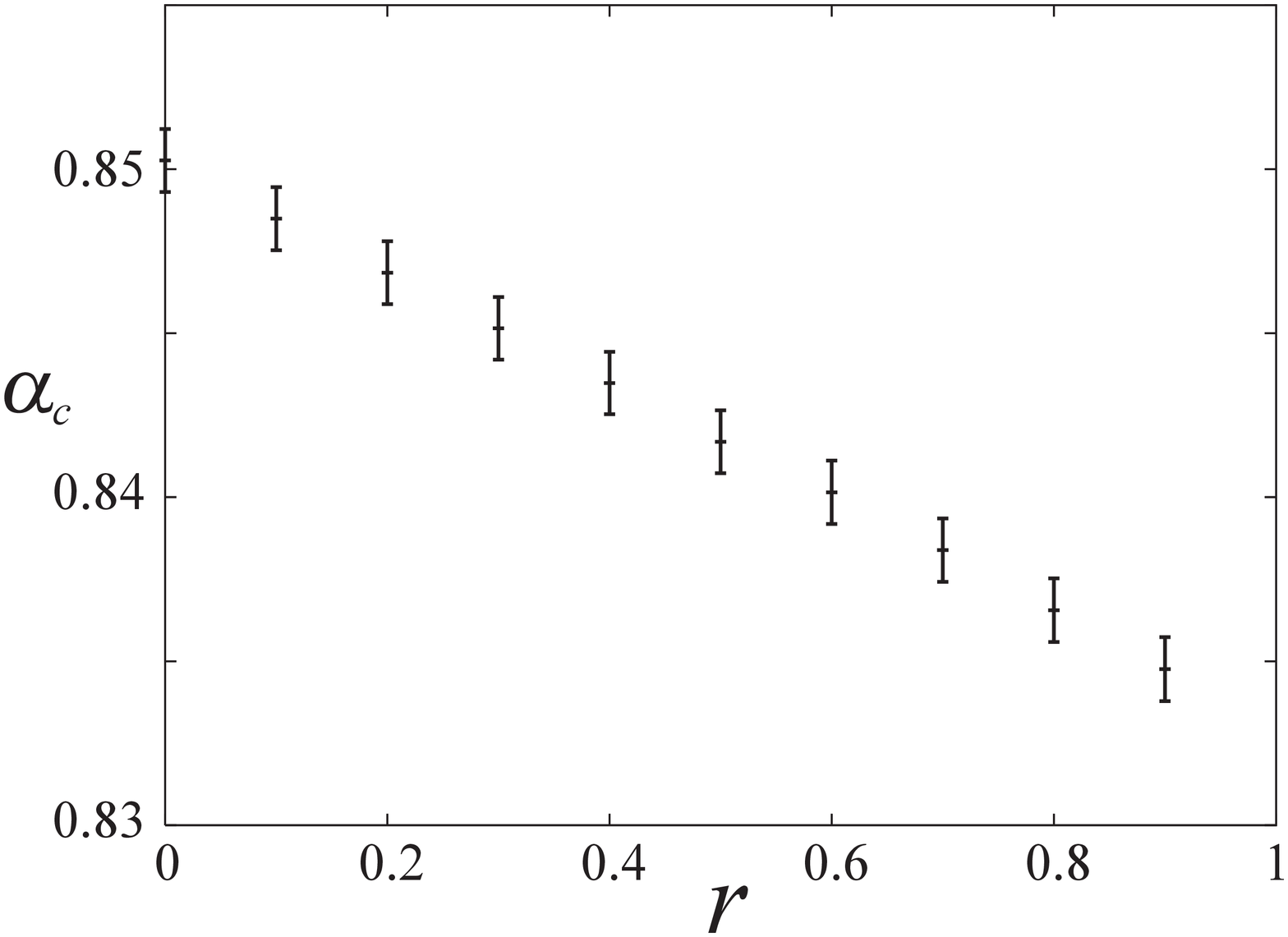}} 
\end{picture}

\caption{Reconstruction limit $\alpha_c$ for signal from SAR(1) as
 function of $\rho$ with $r=0.5$ (top) or $r$ with $\rho=0.5$
 (bottom).
 In the top figure,
 the dependence on $\rho$ (cross) is almost the same as the basic setting
 examined in \cite{KWT} (dotted curve).
}
\label{fig:figure2}
\end{figure}

\begin{figure} 
\begin{picture}(0,115)
 \put(32,-10){\includegraphics[width=0.34\textwidth]{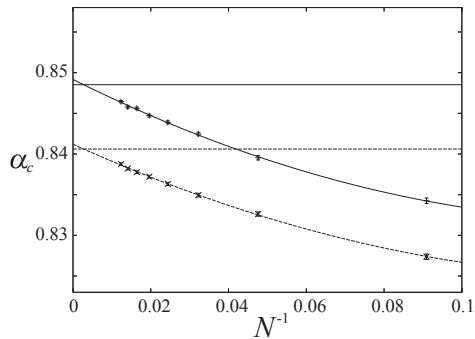}} 
\end{picture} 
\caption{Reconstruction limit $\alpha_c$ for SAR(1) of finite dimension
 $N$ from reconstruction experiment.
 The results for $r=0$ ($+$) and $r=0.5$ ($\times$)
 when $\rho=0.5$ are shown. The curves indicate the result of the
 quadratic function regression for the dependence of (inverse of) input
 signal dimension $N$ (solid for $r=0$ and broken for
 $r=0.5$). The two horizontal lines indicate the results of a replica
 analysis for $r=0$ (solid) and $r=0.5$ (broken), respectively.
}
\label{fig:figure3}
\end{figure}

 We evaluated the reconstruction limit $\alpha_c$ by 
 iteratively solving (\ref{eq:finalresultAR}).
 For numerical stability, we solved the
 equation by converting the coordinates of the variables as $\bm x' =
 \bm S^{-1} \bm x$.  
 We set the length of $\bm x^0$ to $N=2 \times 10^3$
 and took $10^3$ (figure \ref{fig:figure2}) or
 $10^4$ (figure \ref{fig:figure3})
 sample averages for the numerical evaluation of
 $\alpha_c$. Making $N$ much larger is practically difficult due to the
 slow convergence of the iteration under the sample
 fluctuations. However, we judged that the signal length of $N=2 \times
 10^3$ was large enough for the evaluation of $\alpha_c$ because the change in
 $\alpha_c$ evaluated for $N=1 \times 10^3$ was smaller than the value of the
 typical sample fluctuations. 

 The reconstruction limit as a function of $\rho$ and $r$ is depicted in
 figure \ref{fig:figure2}. For a fixed $r$ (top panel), $\alpha_c$
 behaves as a convex upward function of $\rho$ similarly to that for the
 case of the basic setting (dotted curve)
 \cite{KWT}. When comparing this with the
 results from the basic setting of the i.i.d. sparse signals in
 \cite{KWT}, where $\alpha_c = 0.8312 \ldots$ is evaluated for
 $\rho=0.5$, the value of the reconstruction limit for $r=0$, which
 corresponds to cases where there were no time correlations except for
 the pausing, is larger (bottom panel). This implies that the
 reconstruction limit does depend on the types of sparsity and that the
 sparsity of the signal differences is not as useful as that for the
 signals themselves in reducing the data size. With regard to the
 autoregression parameter $r$, a decrease in $\alpha_c$, or the
 equivalent improvement of the reconstruction performance is observed as
 $r$ is increased (bottom panel). This is plausible because the
 correlations generally decrease the information quantity of the
 signals, which in principle makes it possible
 to reduce the data size. 

 To verify the obtained results, we also conducted numerical
 experiments. In the experiments, $\alpha_c$ was numerically assessed as
 follows: In a trial, we first prepared an $N \times N$ random
 compression matrix $\bm F$, and deleted the rows of the matrix
 one-by-one until the signal reconstruction failed. A failure was judged
 when $|\bm x_*-\bm x^0| > 10^{-4}$, where $\bm x_*$ is the
 reconstructed vector, was first satisfied, and the value $Pc=P+1$,
 where $P$ is the number of rows when the reconstruction failure, was
 recorded. We used the convex optimization package for MATLAB developed
 in \cite{GB1,GB2} to search for $\bm x_*$. For each $N$, this trial was
 repeated $10^5$ times, and the typical reconstruction limit for a
 finite $N$, $\alpha_c(N)$, was assessed as
 $\alpha_c(N)=\overline{P_c}/N$, where $\overline{\cdots}$ denotes the
 arithmetic average over the $10^5$ trials. Finally, the critical value
 of $N \to \infty$ was evaluated by
 using the quadratic fitting with respect to $N^{-1}$ to $\alpha_c(N)$. 

 The results are summarized in figure \ref{fig:figure3}, where the
 dependence of $\alpha_c(N)$ on the signal length $N$ is depicted for
 $r=0$ and $r=0.5$ with $\rho=0.5$. A decrease of the reconstruction
 limit $\alpha_c$ (or improvement of reconstruction performance) for a
 larger $r$ is observed as expected from the replica analysis. In order
 to compare this with the reconstruction limit from the replica
 analysis, we also performed a scaling analysis using a quadratic
 function regression and extrapolated the result to $N \rightarrow
 \infty$, which gives $\alpha_c=0.8485(3)$ for $r=0$ and
 $\alpha_c=0.8406(3)$ for $r=0.5$. The reconstruction limits for $N
 \rightarrow \infty$ from the extrapolation are reasonably
 close to the values from the replica analysis
 ($\alpha_c=0.8491(2)$ for $r=0$ and
 $\alpha_c=0.8412(1)$ for $r=0.5$ respectively), 
 considering possible biases which come out due to influences
 of higher order terms of $N^{-1}$ in the data fitting,
 which validates our
 analysis based on the statistical mechanical scheme.

\section{Summary and discussion}
 In summary, we have developed a scheme to assess the typical
 reconstruction limit of compressed sensing problems that are
 defined by the generic signal sources and cost functions under
 the assumption of random measurements. Although the scheme is
 computationally difficult in general, it is still of practical
 utility when the source distribution is computationally feasible
 and the cost function is convex downward. As an example for showing
 the utility, we have taken up the problem of sparse autoregression
 and have examined how $\alpha_c$ depends on two system parameters
 that specify the autoregression process. Our investigation has
 indicated that the sparsity of the signal differences between
 successive times is not as useful as that of the signals themselves
 for compressing the data size. 

 In earlier studies \cite{KWT,DT2009b,DT2010}, the universality of
 $\alpha_c$ has been observed for i.i.d. sparse sources as long as
 the cross correlation matrix $(\bm F)^{\rm T} \bm F$ of the random
 compression matrix $\bm F$ asymptotically obeys a rotationally
 invariant ensemble. The problem of the sparse autoregression of
 vanishing correlation parameter $(r=0)$ can be cast to the cases
 of the i.i.d. sources in which the $(\bm F)^{\rm T} \bm F$ ensemble
 {\em is not} asymptotically rotationally invariant. Our result
 indicates that applying the theoretical results obtained for random
 compression matrices and i.i.d. sources to realistic problems requires
 a certain care because either/both $\bm F$ or/and the original signals
 can contain non-negligible correlations in most real world problems. 
 
 Exploring a more realistic time series modeled by SAR($k$) $(k \ge 2)$,
 two dimensional signals (images)
 is included in our future plan.
 Besides, compressed sensing with noise is also significant for
 application. Its performance can be analyzed by the generalization
 of our formalism, which is also a promising future work.

\acknowledgments
 Support by KAKENHI Nos. 22300003, 22300098,  
 The Mitsubishi Foundation and 
 the JSPS GCOE ``CompView'' is acknowledged (YK).  


\end{document}

%% file: commands.tex
\newcommand{\bx}{\mbox{\boldmath{$x$}}}

\newcommand{\mR}{\mathbb{R}}
\newcommand{\mN}{\mathbb{N}}

\newcommand{\Real}{\mathbb{R}}

\newcommand{\extr}{\mathop{\mathrm{Extr}}}